\documentclass[twocolumn,preprintnumbers,aps]{revtex4}
\usepackage{graphicx}

\newcommand{\be}{\begin{eqnarray}}
\newcommand{\ee}{\end{eqnarray}}

\begin{document}

\title{Weak Equivalence Principle and Propagation of the Wave Function in Quantum Mechanics}

\author{Clovis Jacinto de Matos$^{\rm 1}$}

\affiliation{$^{\rm 1}$European Space Agency, 8-10 rue Mario
Nikis, 75015 Paris, France}

\date{\today}

\preprint{}

\begin{abstract}
The propagation of the wave function of a particle is
characterised by a group and a phase velocity. The group velocity
is associated with the particle's classical velocity, which is
always smaller than the speed of light, and the phase velocity is
associated with the propagation speed of the wave function phase
and is treated as being unphysical, since its value is always
greater than the speed of light. Here we show, using Sciama's
Machian formulation of rest mass energy, that this physical
interpretation, for the group and the phase velocity of the wave
function, is only valid if the weak equivalence principle strictly
holds for the propagating particle, except for the photon. In case
this constraint is released the phase velocity of the wave
function could acquire a physical meaning in quantum condensates.
\end{abstract}

\maketitle

{\sc Introduction ---} A quantum particle is a dual entity with
corpuscular and wave properties. On the one side classically, a
corpuscule is essentially characterized by its position $\vec{r}$,
inertial mass $m_i$, electric charge $q$, linear momentum,
$\vec{p}=m_i\dot{\vec{r}}$, and kinetic energy $E_c=p^2/2m$. On
the other side, a classical wave has the following main
attributes: Wavelength $\lambda$, propagation speed $w$, frequency
$\nu=w/\lambda$, Amplitude $A$, Intensity $I=A^2$, Energy $E$,
linear momentum $\vec{p}$. A corpuscule is a localized entity, in
contrast with a wave which has a certain extension through space.
Quantum mechanics accounts for the dual aspect of quantum entities
by postulating a level of reality more fundamental than the
particle itself. This consists in a complex wave function $\Psi$,
which is \emph{not directly observable}. The value of $\Psi$
associated with a moving particle, at a point with coordinates,
$x$, $y$, and $z$ at time $t$, is related with the probability
$d\pi$ to observe the particle at this particular point.
\begin{equation}
d\pi=|\Psi|^2dx~dy~dz=\Psi\Psi^*~d\tau \label{e1}
\end{equation}
Where $d\tau=dx~dy~dz$ is the infinitesimal volume element in the
neighborhood of the point $(x, y, z)$, and $\Psi^*$ is the complex
conjugate of $\Psi$. The wavelength of the wave function is the so
called de Broglie wavelength, which is inversely proportional to
the particle's linear momentum $p=m_{0i} v/(1-v^2/c^2)^{1/2}$:
\begin{equation}
\lambda=\frac{h}{m_{0i}v}\Big(1-\frac{v^2}{c^2}\Big)^{1/2}\label{e2}
\end{equation}
where $h$ is Planck's constant, and $m_{0i}$ is the particles
proper inertial mass. Assuming that the entire energy of the
particle, $E=E_0/(1-v^2/c^2)^{1/2}$, can be converted into one
single quanta of electromagnetic energy, the frequency of the wave
function $\nu$ is related with the particle's rest mass energy
through the well know Planck's formula:
\begin{equation}
\nu=\frac{E_0}{h \Big(1-\frac{v^2}{c^2}\Big)^{1/2}} \label{e3}
\end{equation}
where $E_0$ is the particle's rest mass energy.

{\sc Machian Interpretation of the Rest Mass Energy ---} Starting
from Mach's principle, which asserts that there is a holistic-type
connection between the local laws of physics and the large scale
properties of the universe, Sciama in \cite{Sciama} introduced the
relation
\begin{equation}
c^2=\frac{2GM}{R}\label{m1}
\end{equation}
where $R$ and $M$ are the radius and the mass of the universe.
Einstein's relationship linking proper energy $E_0$ and proper
mass $m_0$ then takes the form
\begin{equation}
E_0=m_0c^2=\frac{2GMm}{R}\label{m2}
\end{equation}
which can be interpreted as a statement that the proper inertial
energy that is present in any physical particle is due to the
gravitational potential energy of all the matter in the universe
acting on the particle. Therefore the mass $m_0$ appearing in
eq.(\ref{m2}) should be regarded as the proper gravitational mass
$m_{0g}$of the particle.
\begin{equation}
E_0=m_{0g} ~c^2\label{m3}
\end{equation}

{\sc Group and Phase Velocity of the Wave Function ---} To compute
the propagation speed $w$ of the wave function, one starts from:
\begin{equation}
\lambda=\frac{w}{\nu}\label{g1}
\end{equation}
Substituting Eq. (\ref{e2}) and Eq.(\ref{e3}) in Eq.(\ref{g1}) we
obtain,
\begin{equation}
\frac{h}{m_{0i} v}=w \frac{h}{E_0}\label{g2}
\end{equation}
Substituting Eq.(\ref{m3}) in Eq.(\ref{g2}) one gets
\begin{equation}
\frac{m_{0g}}{m_{0i}}=\frac{wv}{c^2}\label{g3}
\end{equation}
For a realistic normalizable wave function forming a wavepacket,
corresponding to the propagation of a free particle, it is easy to
demonstrate \cite{Born}\cite{holland}\cite{Heitler}that $v$ and
$w$ correspond to the group and the wave velocity of the
wavepacket respectively . The former coincides with the velocity
of the particle, the later corresponds to the velocity of
propagation of the phase of the wave function.

{\sc Weak Equivalence Principle and Wave's function Phase velocity
---}
The Weak Equivalence Principle is one of the main foundations of
the theory of general relativity. It means the constancy of the
ratio between the inertial and the gravitational mass $m_i$ and
$m_g$ respectively of a given physical system.
\begin{equation}
\frac{m_g} {m_i}=\iota
\end{equation}
where $\iota$ is a constant. This implies, in classical physics,
that the possible motions in a gravitational field are the same
for different test particles.

Since the weak equivalence principle cannot be demonstrated on a
purely theoretical basis, it can only be justified by experiment.
Thus $\iota$ can only be obtained from experiments. Current
experimental tests of the weak equivalence principle
\cite{Bassler} \cite{Smith}, indicate that the gravitational and
inertial masses of any classical physical system should be equal
to each other
\begin{equation}
m_g/m_i=\iota=1.\label{we2}
\end{equation}
This is observed within a relative accuracy of the
E\"{o}tv\"{o}s-factor, $\eta(A,B)$ less than $5 \times 10^{-13}$.
\begin{equation}
\eta(A,B)= (m_g/m_i)_A - (m_g/m_i)_B<5\times 10^{-13}\label{we3}
\end{equation}
where $A$ and $B$ designate two different bodies, or the same body
at different times.

{\sc Discussion ---} Substituting Eq.(\ref{we2}) in Eq.(\ref{g3})
we deduce the well known relation between the group and phase
velocity of the wave function and the speed of light in vacuum
\cite{Davydov}.
\begin{equation}
wv=c^2\label{d1}
\end{equation}
Since according to the laws of special relativity the group
velocity, i.e. the particle's classical velocity, cannot exceed
the speed of light $c$, the phase velocity is necessarily higher
than the speed of light. This is generally interpreted as
demonstrating the un-physical nature of the phase velocity of the
wave function. \emph{Thus neither the wave function or its wave
velocity can be directly detected and none of them can be
associated with the propagation of information at supra-luminal
speeds}. It is nevertheless worth stressing that this physical
interpretation of the wave function phase velocity is only
possible if the weak equivalence principle holds for the
propagating particle, i.e., if the inertial and gravitational mass
of the particle are exactly equal to each other.

For the sake of completeness, let us investigate what would be the
consequences of requiring that the wave function phase velocity is
physical, i.e. that $w=c$. Substituting this condition in
Eq.(\ref{g3}), we conclude that
\begin{equation}
\frac{m_{0g}}{m_{0i}}=\frac{v}{c}\label{d2}
\end{equation}
Assuming that the weak equivalence principle holds, i.e.
$m_{0g}/m_{0i}=1$, we conclude that Eq.(\ref{d2})leads to $v=c$,
hence it only applies to photons and not to other material
particles. If instead we assume a possible violation of the weak
equivalence principle, in order to allow the phase velocity to be
physical for all types of particles, then Eq.(\ref{d2}) leads us
to understand this symmetry breaking as being equivalent to a
rotation in Minkowsky spacetime between two inertial observers in
relative motion to each other.

In the case of superconductors and superfluids the particles
making the condensate have canonical momentum $\vec p$
proportional to the gradient of the phase of the wave function
$\varphi$.
\begin{equation}
\hbar \nabla \varphi = \vec p \label{da3}
\end{equation}
Assuming an experimental context in which the canonical momentum
$\vec p =m\vec v$, one deduce from Eq.(\ref{da3}), that in
superconductors and superfluids, the group and the phase velocity
of the wave function, $\vec v$ and $\vec w$ respectively, are
equal to each other.
\begin{equation}
\vec v= \frac {\hbar \nabla \varphi}{m}=\vec w \label{db3}
\end{equation}
Substituting Eq.(\ref{db3}) in Eq.(\ref{g3}) one obtains:
\begin{equation}
\frac{m_{0g}}{m_{0i}}=\Big(\frac{v}{c}\Big)^2\label{d3}
\end{equation}
Since in general $v<<c$ in the Earth laboratory, this would mean
that the weak equivalence principle is strongly broken for Cooper
pairs in superconductors and for Helium atoms forming the
superfluid condensate in superfluid Helium. This conclusion is
streamlined with other published work indicating a possible
breaking of the weak equivalence principle for Cooper pairs in
superconductors \cite{clovis1} and for superfluid vortices in
rotating superfluid Helium \cite{clovis2}, resulting from a
breaking of gauge invariance in these physical systems.
Substituting Eq.(\ref{d3}) in Eq.(\ref{we3}) we obtain the
E\"{o}tv\"{o}s factor charactering the breaking of the weak
equivalence principle in quantum condensates, when compared with
the normal state of these materials for which $m_{0g}/m_{0i}=1$.
\begin{equation}
\eta=1-\Big(\frac{v}{c}\Big)^2\label{d4}
\end{equation}

{\sc Conclusion---}In this short note we have demonstrated that
assuming the Machian interpretation of proper energy, which leads
to interpreting the inertial energy content of a particle as being
due to its gravitational interaction with the entire universe, we
can assign the unphysical nature of the phase velocity of the
particle's wave function to the fact that it complies with the
weak equivalence principle, Eq.(\ref{g3}).

Requiring that the wavefunction phase velocity is physical by
propagating at the speed of light, leads to assume either that
this is only possible for the case of the photon, or that the weak
equivalence principle is violated by quantum particles. Since
there is no experimental evidence that this is the case, as
measured in Collela Overhauser Werner (COW)\cite{cow} experiments
with cold neutron interferometry, we conclude that the phase
velocity of the wave function is unphysical except for the photon.

For the special case of superconductors and superfluids, we showed
that the phase velocity of the condensate's wave function would be
physical since being equal to the condenstate's group velocity,
Eq.(\ref{db3}). This leads to consider a possible breaking of the
weak equivalence principle in these materials. This conclusion is
also supported by other research work on the consequences of the
breaking of gauge invariance in superconductors and
superfluids\cite{clovis1} \cite{clovis2}.



\begin{thebibliography}{99}

\bibitem{Sciama} D. W. Sciama, Mon. Not. Roy. Astr. Soc. \textbf{113},
34, (1954)

\bibitem{Born} M. Born, "Atomic Physics", Dover Publication Inc. New
York, pp.89-92, 382-383,  (1969)

\bibitem{holland} P. R. Holland,"The Quantum Theory of Motion",
Cambridge University Press, pp.41-44,(1993)

\bibitem{Heitler} W. Heitler, "\'{E}l\'{e}ments de M\'{e}canique
ondulatoire", Presses Universitaires de France, pp.3-7 (1949)

\bibitem{Davydov} A. S. Davydov, "Quantum Mechanics", Pergamon
International Library, pp. 4-5 (1976)

\bibitem{Bassler}S. Baessler et al., \emph{Phys. Rev. Lett.}
\textbf{83} 3585 (1999)

\bibitem{Smith}G. L. Smith et al., \emph{Phys. Rev.} \textbf{D61}
022001 (2000)

\bibitem{clovis1} C. J. de Matos, "Physical Vacuum in
Superconductors", arXive: 0908.4495 (2009), to appear in J. Sup.
Nov. Mag.

\bibitem{clovis2} C. J. de Matos, "Are Vortices in Rotating
Superfluids Breaking the Weak Equivalence Principle?"
arXiv:0909.2819 (2009), to appear in J. Sup. Nov. Mag.

\bibitem{cow} R. Collela, A. Overhauser, S. A. Werner, \emph{Phys.
Rev. Lett.} \textbf{34}, 1472, (1975)

\end{thebibliography}
\end{document}